\documentclass{appolb}
\usepackage{graphicx}
\newcommand{\ba}{\begin{eqnarray}}
\newcommand{\ea}{\end{eqnarray}}
\newcommand{\bsub}{\begin{subequations}}
\newcommand{\esub}{\end{subequations}}

\def\ket#1{|#1\rangle}

\begin{document}
\title{Partial dynamical symmetry and the\\ 
phonon structure of cadmium isotopes
\thanks{Presented at the Zakopane Conference on Nuclear Physics 
``Extremes of the Nuclear Landscape'', Zakopane, Poland, 
August 26 - September 2, 2018.}%
}
\author{A. Leviatan, N. Gavrielov
\address{Racah Institute of Physics, The Hebrew University, 
Jerusalem 91904, Israel}
}
\maketitle
\begin{abstract}
The phonon structure and spectral properties of states 
in $^{110}$Cd are addressed by
including proton excitations in the phonon basis and exploiting a
partial dynamical symmetry that mixes only certain classes of states
and maintains the vibrational character in the majority of normal states.
\end{abstract}
\PACS{21.60.Fw, 21.10.Re, 21.60.Ev, 27.60.+j}

\vspace{0.5cm}

The cadmium isotopes since long have been considered
as archetypal examples of spherical vibrators, 
manifesting the U(5) dynamical symmetry (DS)~\cite{Iachello87}. 
Recent studies, however, have cast doubt on the validity of this
description~\cite{Garrett08,Garrett12}.
In the present contribution, we address this question
from a symmetry-oriented perspective,
focusing on $^{110}$Cd~\cite{LevGavRamIsa18}.

The U(5)-DS limit of the interacting boson model
(IBM)~\cite{Iachello87}, corresponds to the chain
of nested algebras:
${\rm U(6)\supset U(5)\supset SO(5)\supset SO(3)}$. 
The basis states $\ket{[N],n_d,\tau,n_{\Delta},L}$ 
have quantum numbers which are the labels of irreducible 
representations of the algebras in the chain. 
Here $N$ is the total number of monopole ($s$) and quadrupole ($d$) 
bosons, $n_d$ and $\tau$ are the $d$-boson number and seniority, 
respectively, and $L$ is the angular momentum. 
The multiplicity label $n_{\Delta}$ counts the 
maximum number of $d$-boson triplets coupled to $L\!=\!0$. 
The U(5)-DS Hamiltonian has the form~\cite{Iachello87}
\ba
\hat{H}_{\rm DS} = 
t_1\,\hat{n}_d + t_2\,\hat{n}_{d}^2
+ t_3\,\hat{C}_{{\rm SO(5)}} + t_4\,\hat{C}_{{\rm SO(3)}} ~,
\label{H-DS}
\ea
where $\hat{C}_{\rm G}$ is a Casimir operator of G, and 
$\hat{n}_d\!=\!\sum_{m}d^{\dag}_md_m\!=\!\hat{C}_{{\rm U(5)}}$. 
$\hat{H}_{\rm DS}$ is completely 
solvable for {\it any} choice of parameters $t_i$, with eigenstates
$\ket{[N],n_d,\tau,n_{\Delta},L}$ and energies 
$E_{\rm DS} \!=\!t_1 n_d + t_2 n_{d}^2 + t_3 \tau(\tau+3) + t_4 L(L+1)$. 
A typical U(5)-DS spectrum exhibits
$n_d$-multiplets of a spherical vibrator, 
with enhanced connecting ($n_d+1\!\rightarrow\! n_d$) $E2$ transitions.
\begin{figure}[t]\centering
\begin{minipage}[t]{0.5\linewidth}
\includegraphics[width=\linewidth]{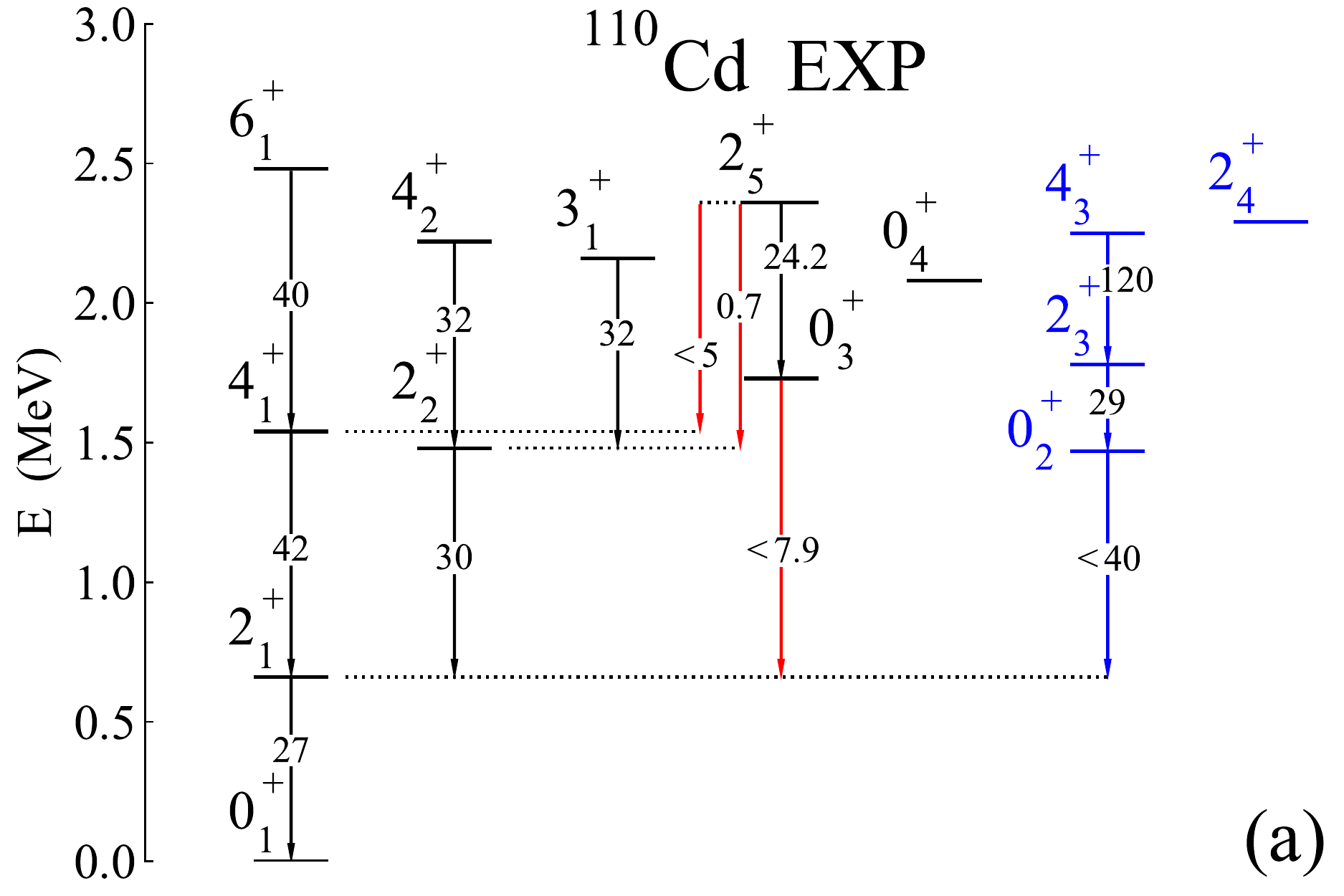}
\end{minipage}\hfill
\begin{minipage}[t]{0.5\linewidth}
\includegraphics[width=\linewidth]{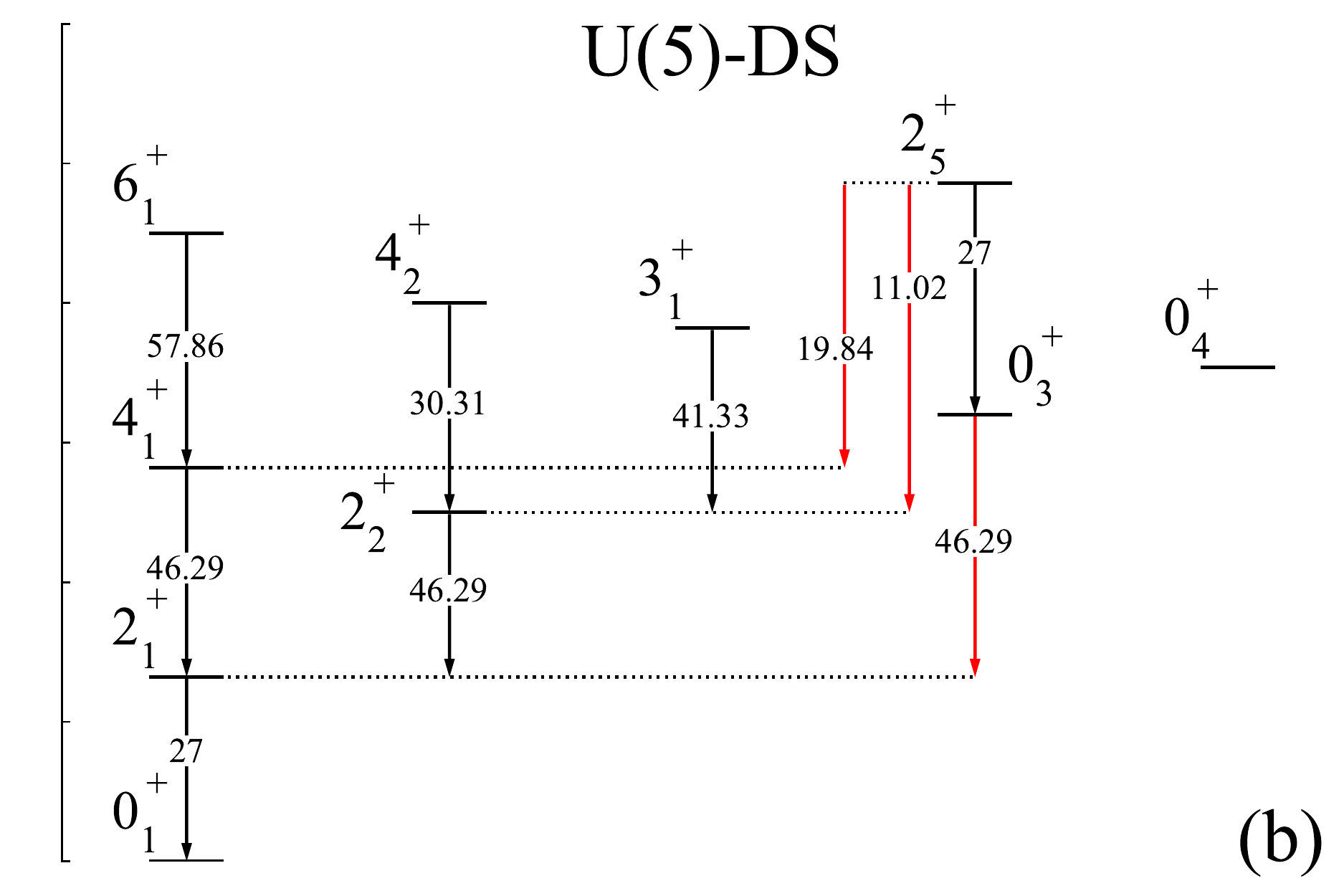}
\end{minipage}
\vspace{12pt}
\begin{minipage}[t]{0.5\linewidth}
\includegraphics[width=\linewidth]{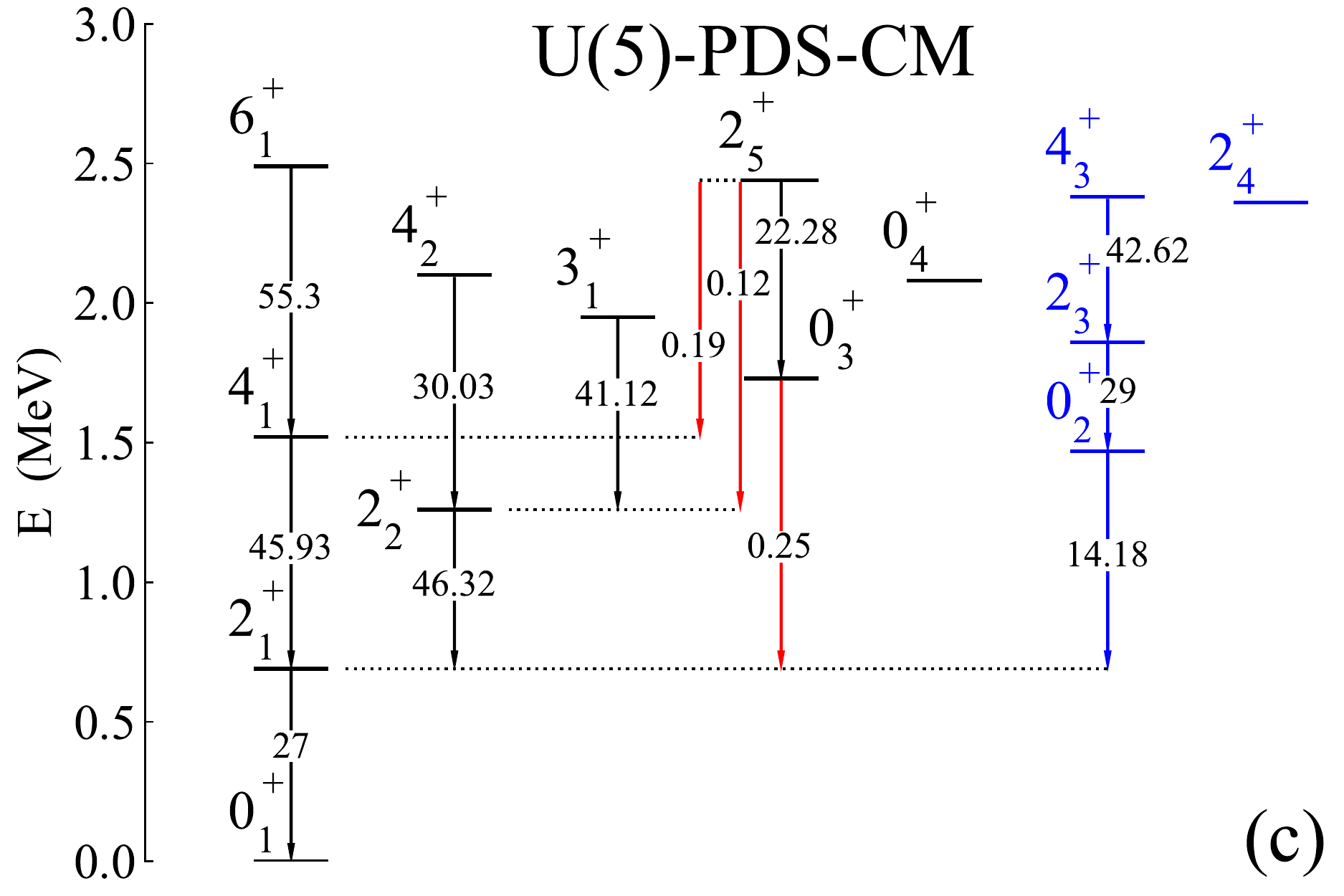}
\end{minipage}\hfill
\begin{minipage}[t]{0.5\linewidth}
\includegraphics[width=\linewidth]{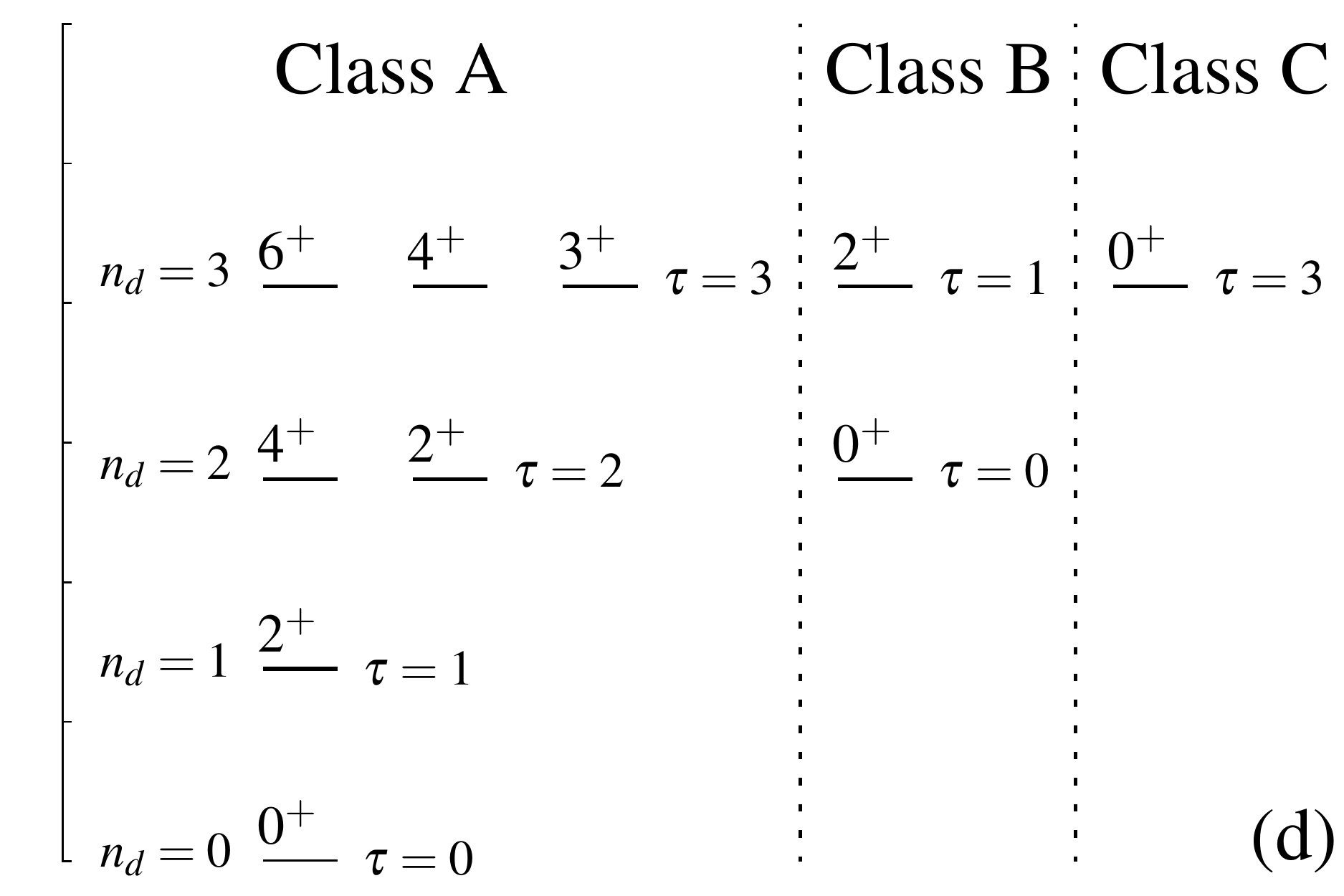}
\end{minipage}\hfill
\caption{\small
(a)~Experimental spectrum and representative 
$E2$ rates~\cite{Garrett12,NDS12} (in W.u.)  of normal 
and intruder levels ($0^{+}_2,\,2^{+}_3,\,4^{+}_3,\,2^{+}_4$)
in $^{110}$Cd. 
(b)~Calculated U(5)-DS spectrum obtained from 
$\hat{H}_{\rm DS}$~(\ref{H-DS}) with parameters 
$t_1\!=\!715.75,\, t_2\!=\!-t_3\!=\!42.10,\, 
t_4\!=\!11.38$ keV and $N\!=\!7$. 
(c)~Calculated U(5)-PDS-CM spectrum, obtained from 
$\hat{H}$~(\ref{Hfull}) with parameters 
$t_1\!=\!767.83,\, t_2\!=\!-t_3\!=\!73.62,\, 
t_4\!=\!18.47,\,r_0\!=\!2.15,\, e_0\!=\!-6.92,\,
\kappa\!=\!-72.73,\,\Delta\!=\!9978.86,\,\alpha\!=\!-42.78$ 
keV and $N\!=\!7\,(9)$ in the normal (intruder) sector.
(d) Classes of low-lying U(5)-DS states.} 
\end{figure}  
\begin{table}[]
\begin{center}
\caption{\label{TableI}
\small
Absolute (relative in square brackets) 
$B(E2)$ values in W.u.
for $E2$ transitions from normal levels in $^{110}$Cd.
The experimental (EXP) values are taken from~\cite{Garrett12,NDS12}. 
The U(5)-DS [U(5)-PDS-CM] values are obtained for an $E2$ operator 
$e_{B}\,\hat{Q}$
[$e_{B}^{(N)}\hat{Q}^{(N)} + e_{B}^{(N+2)}\,\hat{Q}^{(N+2)}$] 
with $e_{B}\!=\!1.964$ [$e_{B}^{(N)}\!=\!1.956$ and 
$e_{B}^{(N+2)}\!=\!1.195$] $(\rm W.u.)^{1/2}$, 
where $\hat{Q} = d^{\dag}s + s^{\dag}\tilde{d}$ and $\hat{Q}^{(N)}$ 
denotes its projection onto the $[N]$ boson space. 
In both calculations the 
boson effective charges were fixed by the empirical 
$2^{+}_1\rightarrow 0^{+}_1$ rate. 
Intruder states $0^{+}_{2;i}\,2^{+}_{3;i},\,4^{+}_{3;i},\,2^{+}_{4;i}$,
are marked by a subscript~$i$.}
\begin{tabular}{lllcc}
\hline
$L_i$ & $ L_f$ & EXP & U(5)-DS & U(5)-PDS-CM \\[1pt]
\hline
$2^+_{1}$ & $0^+_{1}$ & 27.0 (8) & 27.00 & 27.00 \\[1pt]

$4^+_{1}$ & $2^+_{1}$ & 42 (9)   & 46.29 & 45.93 \\
$2^+_{2}$ & $2^+_{1}$ & 30 (5); 19 (4)$^a$
   & 46.29 & 46.32 \\
         & $0^+_{1}$ & 1.35 (20); 
0.68 (14)$^a$ & 0.00 & 0.00 \\
$0^+_{3}$ & $2^+_{2}$ & $<$ 1680$^a$  & 0.00 & 55.95 \\
         & $2^+_{1}$ & $<$ 7.9$^a$ & 46.29 & 0.25  \\
$6^+_{1}$ & $4^+_{1}$ & 40 (30); 62 (18)$^a$         & 57.86 & 55.30 \\[1pt]
         & $4^+_{2}$ & $<$ 5$^a$ & 0.00  & 0.00 \\[1pt]
         & $4^+_{3;i}$ & 14 (10); 36 (11)$^a$       &     & 2.39 \\         
$4^+_{2}$ & $4^+_{1}$ & 12$^{+4}_{-6}$; $^a$10.7$^{+4.9}_{-4.8}$  & 27.55
& 27.45 \\[1pt]
         & $2^+_{2}$ & 32$^{+10}_{-14}$; 22 (10)$^a$  & 30.31 & 30.03 \\[1pt]
& $2^+_{1}$ & 0.20$^{+0.06}_{-0.09}$; 0.14 (6)$^a$    & 0.00
& 0.00  \\[1pt]
         & $2^+_{3;i}$ & $<$ 0.5$^a$ &        & 0.005 \\[1pt]                
$3^+_{1}$ & $4^+_{1}$ & 5.9$^{+1.8}_{-4.6}$; $^a$2.4$^{+0.9}_{-0.8}$ 
& 16.53 & 16.48 \\[1pt]
         & $2^+_{2}$ & 32$^{+8}_{-24}$; 22.7 (69)$^a$ & 41.33 & 41.12 \\[1pt]
	 & $ 2^+_{1}$ & 1.1$^{+0.3}_{-0.8}$; 0.85 (25)$^a$   & 0.00  & 0.00 \\[1pt]
                     & $ 2^+_{3;i}$ & $<$ 5$^a$  &        & 0.012 \\[1pt]
$ 0^+_{4}$ & $2^+_{2}$ & [$<$ 0.65$^a$]  & 57.86 & 1.24 \\
                     & $2^+_{1}$ & [0.010$^a$]  & 0.00  & 31.76 \\
                     & $2^+_{3;i}$ & [100$^a$]  &        & 16.32 \\
$2^+_{5}$ & $0^+_{3}$ & 24.2 (22)$^a$  & 27.00 & 22.28 \\
                    & $4^+_{1}$ & $<$5$^a$  & 19.84 & 0.19 \\
                    & $2^+_{2}$ &$^a$0.7$^{+0.5}_{-0.6}$ & 11.02 & 0.12 \\[1pt]
         & $2^+_{1}$ & 2.8$^{+0.6}_{-1.0}$                  & 0.00  & 0.00 \\[1pt]
	 & $ 2^+_{3;i}$ & $<$ 5$^a$  &       & 0.002 \\
	 & 
$ 0^+_{2;i}$ & $<$ 1.9$^a$ &       & 0.20 \\[2pt]
\hline
\end{tabular}
\end{center}
\label{tab:transitions-normal}
\vspace{-0.2cm}
$^a$ {\small From Ref.~\cite{Garrett12}}.
\end{table}
\begin{table}[]
\begin{center}
%\vspace{-1.1cm}
\caption{\label{TableII}
\small
$B(E2)$ values (in W.u.)
for $E2$ transitions from intruder levels in $^{110}$Cd.
Notation and relevant information on the observables shown, 
are as in Table~1.}
\begin{tabular}{lllc}
\hline
$L_i$ $\;\;$ & $ L_f$ $\;\;$ & EXP $\;\;$ & U(5)-PDS-CM \\[1pt]
\hline
$0^+_{2;i}$ & $2^+_{1}$ & 
$<$ 40$^a$     & 14.18 \\[2pt]
$ 2^+_{3;i}$ & $0^+_{2;i}$ & 29 (5)$^a$           & 29.00 \\[1pt]
	    & $0^+_{1}$  & 0.31$^{+0.08}_{-0.12}$; 0.28 (4)$^a$  & 0.08 \\[1pt]
& $2^+_{1}$ & 0.7$^{+0.3}_{-0.4}$; $^a$0.32$^{+0.10}_{-0.14}$
& 0.00 \\[1pt]
            & $2^+_{2}$ & $<$ 8$^a$              & 0.96 \\[1pt]
$ 2^+_{4;i}$ & $2^+_{1}$ & 0.019$^{+0.020}_{-0.019}$ & 0.10 \\[1pt]
$ 4^+_{3;i}$ & $2^+_{1}$ & 0.22$^{+0.09}_{-0.19}$; 0.14 (4)$^a$  & 0.49 \\[1pt]
            & $2^+_{2}$ & 2.2$^{+1.4}_{-2.2}$; 1.2(4)$^a$ & 0.00 \\[1pt]
            & $2^+_{3;i}$ & 120$^{+50}_{-110}$; 115 (35)$^a$   & 42.62 \\[1pt]
& $4^+_{1}$   & 2.6$^{+1.6}_{-2.6}$; $^a$1.8$^{+1.0}_{-1.5}$
& 0.00 \\[2pt]
\hline
\end{tabular}
\end{center}
\label{tab:transitions-intruder}
\end{table}	

The empirical spectrum of $^{110}$Cd, shown in Fig.~1(a), 
consists of both normal and intruder levels, the latter 
based on 2p-4h proton excitations across the $Z\!=\!50$ closed shell. 
Experimentally known $E2$ rates are listed in Tables~1-2.
A comparison of the calculated spectrum [Fig.~(1b)] 
and $B(E2)$ values [Table~1], 
obtained from $\hat{H}_{\rm DS}$~(\ref{H-DS}),
demonstrates that most normal states have good spherical-vibrator
properties, and conform well with the properties of U(5)-DS. 
However, the measured 
rates for $E2$ decays from the non-yrast states, $0^{+}_3\,(n_d\!=\!2)$ 
and $[0^{+}_4,\, 2^{+}_5\,(n_d\!=\!3)]$, 
reveal marked deviations from this behavior. 
In particular, 
$B(E2;\,0^{+}_3\!\rightarrow\! 2^{+}_1) \!<\! 7.9$, 
$B(E2;\,2^{+}_5\!\rightarrow\! 4^{+}_1) \!<\! 5$,
$B(E2;\,2^{+}_5\!\rightarrow\! 2^{+}_2) \!=\! 0.7^{+0.5}_{-0.6}$ W.u.,  
are extremely small compared to the 
U(5)-DS values: $46.29$, $19.84$, $11.02$ W.u., respectively. 
Absolute $B(E2)$ values for transitions from the $0^{+}_4$ state 
are not known, but its branching ratio to $2^{+}_2$ is small.

Attempts to explain the above deviations in terms of 
mixing between the normal spherical [U(5)-like] states and 
intruder deformed [SO(6)-like] states have been shown to be 
unsatisfactory~\cite{Garrett08,Garrett12}. 
This has led to the conclusion that
the normal-intruder strong-mixing scenario needs to be rejected,
and have raised serious questions on the appropriateness 
of the multi-phonon interpretation~\cite{Garrett08,Garrett12}. 
In what follows, we consider a possible explanation for the 
``Cd problem'', based on U(5) partial dynamical symmetry (PDS). 
The latter corresponds to a situation in which the U(5)-DS
is obeyed by only a subset of states and is broken in other
states~\cite{Leviatan11}. Similar PDS-based approaches have been 
implemented in nuclear spectroscopy, in conjunction with the 
SU(3)-DS~\cite{Leviatan96,LevSin99,levramisa13} and 
SO(6)-DS~\cite{Ramos09,LevGav17} chains of the IBM.

As depicted in Fig.~1(d), the lowest spherical-vibrator levels comprise
three classes of states. Specifically, 
Class A: $n_d\!=\!\tau\!=\!0,1,2,3$ $(n_{\Delta} \!=\! 0)$;
Class B: $n_d \!=\! \tau+2\!=\!2,3$ $(n_{\Delta} \!=\! 0)$;
Class~C: $n_d \!=\! \tau \!=\! 3$ $(n_{\Delta} \!=\! 1)$. 
In the U(5)-DS calculation of Fig~1(b), applicable to normal states only, 
the ``problematic'' states 
$[0^{+}_3\,(n_d\!=\!2)$ and $2^{+}_5\,(n_d\!=\!3)]$  
belong to class~B, and $0^{+}_4\, (n_d\!=\!3)$ belongs to class~C.
The remaining ``good'' spherical-vibrator states 
$[0^{+}_1\,(n_d\!=\!0);\,2^{+}_1\,(n_d\!=\!1);\,
4^{+}_1,2^{+}_2\,(n_d\!=\!2);\,6^{+}_1,4^{+}_2,3^{+}_1\,(n_d\!=\!3)]$ 
belong to class~A. As mentioned, the spherical-vibrator interpretation 
is valid for most normal states in Fig.~1(a), but not all. 
We are thus confronted with a situation in which some states 
in the spectrum (assigned to class~A) obey the predictions of U(5)-DS, 
while other states (assigned to classes B and C) do not. 
These empirical findings signal the presence of U(5)-PDS.

The construction of an Hamiltonian with U(5)-PDS follows the general
algorithm~\cite{Leviatan11} and leads to the form:
\ba
\hat{H}_{\rm PDS} &=& \hat{H}_{\rm DS} +
r_0\,G^{\dag}_{0}G_{0}
+ e_{0}\,\left (G^{\dag}_0 K_0 + K^{\dag}_{0}G_0 \right ) ~,
\label{H-PDS}
\ea
where $\textstyle{G^{\dag}_{0} \!=\! [(d^\dag d^\dag)^{(2)} d^\dag]^{(0)}}$,
$K^{\dag}_{0} \!=\! s^{\dag}(d^{\dag} d^{\dag})^{(0)}$. 
The last two terms in Eq.~(\ref{H-PDS}) annihilate the states
$\ket{[N], n_d=\tau, \tau, n_{\Delta}=0, L}$
with $L\!=\!\tau,\tau+1,\ldots,2\tau-2,2\tau$. 
These states, which include those of class~A,
form a subset of U(5) basis states, hence remain solvable eigenstates 
of $\hat{H}_{\rm PDS}$~(\ref{H-PDS}) with good U(5) symmetry. 
It should be noted that while $\hat{H}_{\rm DS}$~(\ref{H-DS}) is diagonal 
in the U(5)-DS chain, the $r_0$ and $e_0$ terms can
connect states with different $n_d$ and/or $\tau$. 
Accordingly, the remaining eigenstates of $\hat{H}_{\rm PDS}$~(\ref{H-PDS}), 
in particular those of classes B and C, 
are mixed with respect to U(5) and SO(5). 
The U(5)-DS is thus preserved in a subset of eigenstates, 
for any choice of parameters in $\hat{H}_{\rm PDS}$, but is broken in others.
By definition, $\hat{H}_{\rm PDS}$ exhibits U(5)-PDS. 
\begin{figure}[t]
\centerline{%
\includegraphics[width=10.5cm]{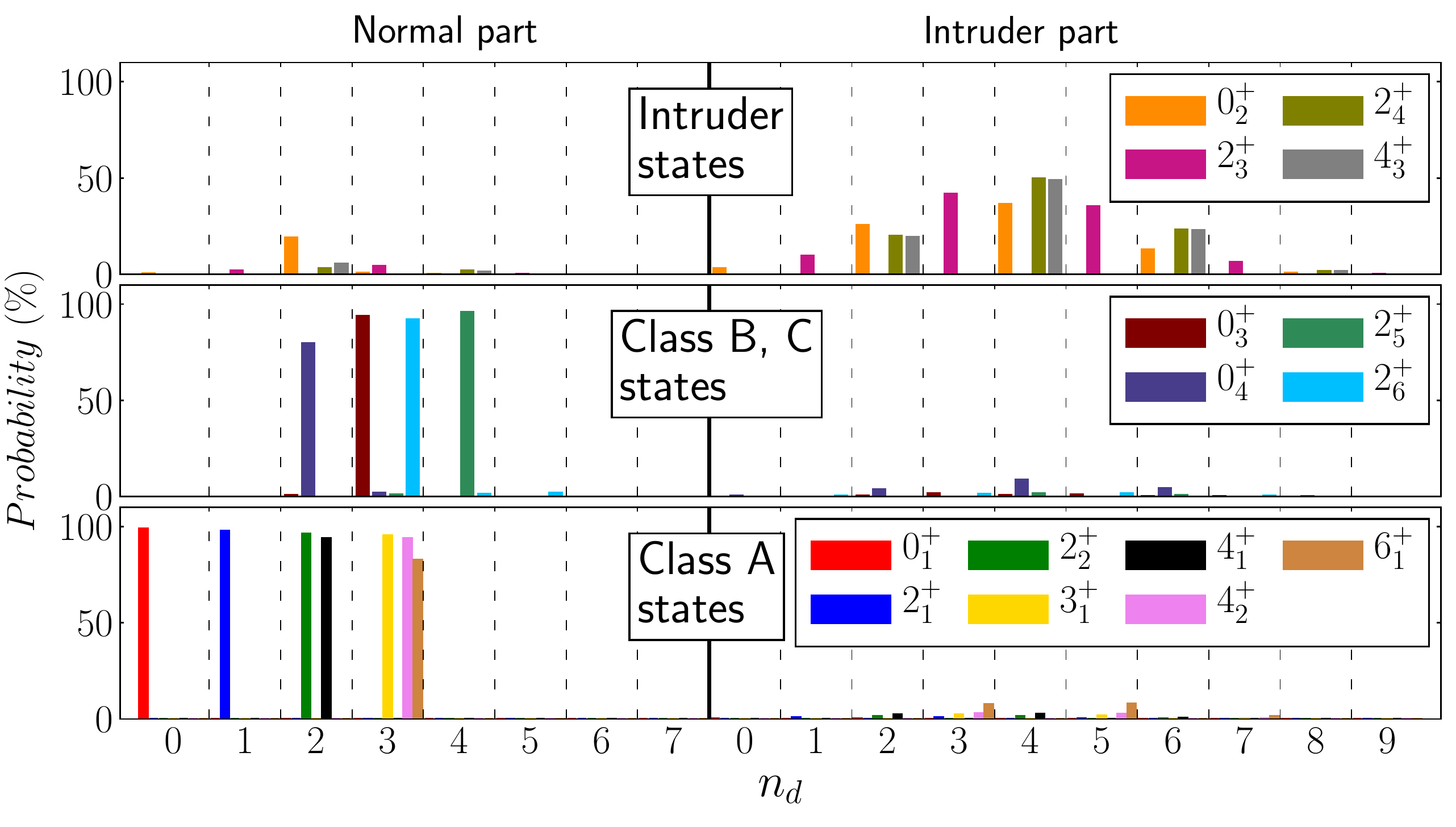}}
\caption{\small
U(5) $n_d$-decomposition of Class~A (lower panel),
Classes B, C (middle panel) and intruder (upper panel) states
in the U(5)-PDS-CM calculation of Fig.~1(c). In each panel,
the left (right) hand side displays the $n_d$-probabilities for the
normal (intruder) components of the total wave function
$\ket{\Psi} \!=\! a\ket{\Psi_{n}^{(N)}} + b\ket{\Psi_{i}^{(N+2)}}$.}
\end{figure}

The combined effect of normal and intruder states, can be studied 
within the interacting boson model with configuration mixing
(IBM-CM)~\cite{DuvBar81}. 
The Hamiltonian for the two configurations has the
form~\cite{LevGavRamIsa18},
\ba
\hat{H} = \hat{H}_{\rm PDS}^{(N)} + \hat{H}_{\rm intrud}^{(N+2)} 
+ \hat {V}_{\rm mix} ~.
\label{Hfull}
\ea
For $^{110}$Cd, 
the Hamiltonian in the normal sector is taken to be $\hat{H}_{\rm PDS}$ of 
Eq.~(\ref{H-PDS}), projected onto a space of $N\!=\!7$ bosons. 
The SO(6)-type of Hamiltonian in the intruder sector is 
$\hat H_{\rm intrud} \!=\! \kappa \hat{Q}\cdot \hat{Q} + \Delta$, 
projected onto a space of $N\!=\!9$ bosons. 
$\hat V_{\rm mix} \!=\!  
\alpha [(s^{\dagger})^{2} + (d^{\dagger}d^{\dagger})^{(0)}]
+ {\rm H.c.}$ is a mixing term between the two spaces. 
In general, an eigenstate of $\hat{H}$,
$\ket{\Psi} \!=\! a\ket{\Psi_{n}^{(N)}} + b\ket{\Psi_{i}^{(N+2)}}$, 
involves a mixture of normal~($n$) and intruder~($i$) components 
with $N$  and $N\!+\!2$ bosons, respectively. 

As seen in Fig.~1(c) and Tables~1-2,
the IBM-PDS-CM calculation provides a good description 
of the empirical data in $^{110}$Cd. 
The U(5) decomposition of the resulting eigenstates are shown in Fig.~2.
The normal states of class~A retain 
good U(5) symmetry to a good approximation. 
Their $\ket{\Psi_n^{(N)}}$ part involves a single $n_d$-component. 
The mixing with the intruder states is weak 
(small $b^2$) of order a few percent.
The high degree of purity is reflected 
in the calculated $B(E2)$ values for transitions between class A states, 
which are very similar to those 
of U(5)-DS. In contrast, the structure of the non-yrast states 
assigned originally to classes~B and C, whose decay properties show 
marked deviations from the U(5)-DS limit, changes dramatically.
Specifically, the $0^{+}_3$ and 
$0^{+}_4$ states, which in the U(5)-DS classification are members of the 
two-phonon triplet and three-phonon quintuplet, 
interchange their character, and the U(5) decomposition of their 
$\ket{\Psi_n^{(N)}}$ parts peaks at $n_d\!=\!3$ and $n_d\!=\!2$, respectively. 
Similarly, the $2^{+}_5$ state, originally 
a member of the three-phonon quintuplet, 
its $\ket{\Psi_n^{(N)}}$ part exhibits a peak at $n_d\!=\!4$.
The calculated 
$B(E2;\,0^{+}_3\!\rightarrow\! 2^{+}_1) \!=\! 0.25$, 
$B(E2;\,2^{+}_5\!\rightarrow\! 4^{+}_1) \!=\! 0.19$ and 
$B(E2;\,2^{+}_5\!\rightarrow\! 2^{+}_2) \!=\! 0.12$ W.u.,
are consistent with the measured upper limits:
$7.9,\,5$ and $0.7^{+0.5}_{-0.6}$ W.u., respectively.
The vibrational interpretation is thus maintained in the
majority of low-lying normal states in $^{110}$Cd.

This work was done in collaboration with J.E.~Garc\'\i a-Ramos (Huelva)
and P.~Van~Isacker (GANIL) and is supported by the Israel Science 
Foundation (Grant 586/16).

\end{document}